\documentclass[10pt,twocolumn,letterpaper]{article}

\usepackage[pagenumbers]{arXiv} 

\usepackage{graphicx}
\usepackage{amsmath}
\usepackage{amssymb}
\usepackage{booktabs}
\usepackage[normalem]{ulem}
\usepackage{numprint}
\usepackage{mathtools, mathdots}
\usepackage{subcaption}
\usepackage{xcolor}
\usepackage{xfrac}
\usepackage[ruled,vlined]{algorithm2e}
\usepackage{algorithmic}
\usepackage[stable]{footmisc}

\usepackage[pagebackref,breaklinks,colorlinks]{hyperref}

\newcommand\narrowdots{\hbox to 0.8em{\scriptsize $\cdot$\hss$\cdot$\hss$\cdot$}}

\makeatletter
\@namedef{ver@everyshi.sty}{}
\makeatother

\usepackage{tikz}\usepackage{tikz, pgfplots}
\usetikzlibrary{calc}
\usetikzlibrary{math}
\usetikzlibrary{shapes}
\usetikzlibrary{patterns}
\usetikzlibrary{backgrounds}
\usetikzlibrary{positioning}
\usetikzlibrary{decorations.pathmorphing}

\usepgfplotslibrary{fillbetween}

\definecolor{plotblack}{HTML}{252422}
\definecolor{plotgrey}{HTML}{f4efef}

\definecolor{plotblue}{HTML}{045275}
\definecolor{plotteal}{HTML}{089099}
\definecolor{plotgreen}{HTML}{7ccba2}
\definecolor{plotyellow}{HTML}{ffc61e}  
\definecolor{plotorange}{HTML}{f0746e}
\definecolor{plotred}{HTML}{dc3977}
\definecolor{plotpurple}{HTML}{7c1d6f}

\colorlet{ploter}{plotyellow}
\colorlet{plotcl}{plotorange}
\colorlet{plotsbm}{plotred}
\colorlet{plotgraphrnn}{plotpurple}
\colorlet{plotverg}{plotteal}
\colorlet{plotdyverg}{plotgreen}

\colorlet{plotemaildnc}{plotyellow}
\colorlet{plotemaileucore}{plotorange}
\colorlet{plotcoauthdblp}{plotred}
\colorlet{plotfacebooklinks}{plotpurple}

\tikzmath{%
    \figscale = 1.75;
    \nodesize = 0.06cm; 
    \nodespacing = 1.5pt;
    \radbig = 360/5;
    \radtri = 360/3;
    \rottri = 360/6;
    \radpent = 360/5;
    \rotpent = 360/10;
    \arccurve = 0.5cm;
    \arcstraight = 0.0cm;
}

\tikzset{snake it/.style={
    decoration={snake, 
        amplitude = 0.2mm,
        segment length = 1mm,
        post length=1mm},decorate}
}

\tikzstyle{blank} = [outer sep=8*\nodespacing, inner sep=\nodesize]
\tikzstyle{bdeg} = [outer sep=\nodespacing, inner sep=\nodesize]

\tikzstyle{tnode} = [circle, draw=plotgrey, outer sep=0cm, inner sep=\nodesize/4, fill opacity=1, draw opacity=1, text opacity=1, fill=white]
\tikzstyle{taddnode} = [tnode, fill=plotteal, fill opacity=0.25]
\tikzstyle{tdelnode} = [tnode, fill=plotorange, fill opacity=0.25]

\tikzstyle{lhs} = [very thick, rectangle, rounded corners=0.5mm, draw=plotblack, fill=white, outer sep=4*\nodespacing, inner sep=\nodesize,  fill opacity=1, draw opacity=1, text opacity=1]
\tikzstyle{addlhs} = [lhs, draw=plotteal, fill=plotteal, fill opacity=0]
\tikzstyle{dellhs} = [lhs, draw=plotorange, fill=plotorange, fill opacity=0]

\tikzstyle{nts} = [very thick, rectangle, rounded corners=0.5mm, draw=plotblack, fill=white, outer sep=\nodespacing, inner sep=\nodesize, fill opacity=1, draw opacity=1, text opacity=1]
\tikzstyle{addnts} = [nts, draw=plotteal, fill=plotteal, fill opacity=0]
\tikzstyle{delnts} = [nts, draw=plotorange, fill=plotorange, fill opacity=0]

\tikzstyle{node} = [very thick, circle, draw=plotblack, fill=plotblack, outer sep=\nodespacing, inner sep=\nodesize, text opacity=1]  
\tikzstyle{addnode} = [node, draw=plotteal, fill=plotteal, draw opacity=1]
\tikzstyle{delnode} = [node, draw=plotorange, fill=plotorange, opacity=1]

\tikzstyle{thinedge} = [thick, draw=plotblack, opacity=1]
\tikzstyle{edge} = [very thick, draw=plotblack, opacity=1]
\tikzstyle{addedge} = [edge, densely dotted, draw=plotteal]
\tikzstyle{frontieredge} = [edge, densely dash dot, draw=plotteal]
\tikzstyle{deledge} = [edge, densely dotted, draw=plotorange, decorate, decoration={snake, amplitude=0.25mm, segment length=2.7mm, post length=0.0mm, pre length=0.0mm}]

\tikzstyle{bedge} = [edge, opacity=1]
\tikzstyle{addbedge} = [bedge, densely dotted, draw=plotteal]
\tikzstyle{delbedge} = [bedge, densely dotted, draw=plotorange, decorate, decoration={snake, amplitude=0.25mm, segment length=2.7mm, post length=0.0mm, pre length=0.0mm}]

\tikzstyle{fedge} = [very thick, draw=plotblack, rounded corners=0.05cm, opacity=1]
\tikzstyle{addfedge} = [fedge, densely dotted, draw=plotteal]
\tikzstyle{delfedge} = [very thick, densely dotted, draw=plotorange, decorate, decoration={snake, amplitude=0.25mm, segment length=2.7mm, post length=0.0mm, pre length=0.0mm}]

\tikzstyle{cone} = [draw=plotgrey, shorten <= -2pt]
\tikzstyle{addcone} = [draw=plotgrey, shorten <= -2pt]

\usetikzlibrary{pgfplots.groupplots}
\usetikzlibrary{shapes.geometric}
\usetikzlibrary{positioning,fit,shapes.geometric,backgrounds, calc}
\usetikzlibrary{arrows,decorations.markings}
\usetikzlibrary{shapes.arrows}
\usetikzlibrary{patterns}
\tikzset{textnode/.style={inner sep=0pt,outer sep=0,execute at begin node={\strut}}}
\tikzstyle{state} = [textnode,circle, draw, inner sep=0pt, outer sep=0]
\usepgfplotslibrary{groupplots}
                    
\pgfplotsset{every axis/.append style={
                    xlabel={$x$},          
                    ylabel={$y$},          
                    ylabel near ticks,
                    xlabel near ticks,
                    legend cell align={left},
                    },
                    legend image code/.code={
                    \draw[mark repeat=2,mark phase=2]
                        plot coordinates {
                        (0cm,0cm)
                        (0.15cm,0cm)        
                        (0.3cm,0cm)         
                        };%
                    }
                    }
\pgfplotsset{compat=newest}

\hypersetup{
    colorlinks,
    linkcolor={plotred},
    citecolor={plotblue},
    urlcolor={plotteal}
}

\usepackage[capitalize]{cleveref}
\crefname{section}{Sec.}{Secs.}
\Crefname{section}{Section}{Sections}
\Crefname{table}{Table}{Tables}
\crefname{table}{Tab.}{Tabs.}

\def\cvprPaperID{1} 
\def\confName{CVPR}
\def\confYear{2023}

\makeatletter
\apptocmd\@maketitle{{%
    \centering
    \scalebox{2}{\begin{tikzpicture}
    \begin{scope}[scale=1]  
        \node [node, draw=plotblack, fill=plotgrey] at (0, 0) (ego) {$v$};

        \node [node, draw=plotred, fill=plotred] at (0, 1) (b) {};
        \node [node, draw=plotorange, fill=plotorange] at (-0.7071067, 0.7071067) (c) {};
        \node [node, draw=plotyellow, fill=plotyellow] at (-1, 0) (d) {};
        \node [node, draw=plotteal, fill=plotteal] at (1, 0) (e) {};
        \node [node, draw=plotblue, fill=plotblue] at (0.7071067, -0.7071067) (f) {};
    \end{scope}
    \begin{scope}[scale=1] 
        \draw [edge] (ego) -- (b);
        \draw [edge] (ego) -- (c);
        \draw [edge] (ego) -- (d);
        \draw [edge] (ego) -- (e);
        \draw [edge] (ego) -- (f);

        \draw [edge] (e) -- (f);
    \end{scope}
\end{tikzpicture}
}
    ~~~~~~~~~~~~~~~~
    \scalebox{1}{\tikzmath{%
    \colone = 0;
    \coltwo = 5;
    \yoffset = 1;
    \yshiftrulefour = -0.75;
}

\begin{tikzpicture}
    \coordinate (left1l) at (\colone, -1 * \yoffset);
    \coordinate (right1l) at (\colone + 1.5, -1 * \yoffset);
    \coordinate (label1r) at (\coltwo - 3/4, -1 * \yoffset);
    \coordinate (left1r) at (\coltwo, -1 * \yoffset);
    \coordinate (right1r) at (\coltwo + 1.5, -1 * \yoffset);
    \coordinate (arrow1) at (\colone + 0.75, -0.75 * \yoffset);

    \coordinate (left2l) at (\colone, -2.1 * \yoffset);
    \coordinate (right2l) at (\colone + 1.5, -2.1 * \yoffset);
    \coordinate (label2r) at (\coltwo - 3/4, -2.1 * \yoffset);
    \coordinate (left2r) at (\coltwo, -2.1 * \yoffset);
    \coordinate (right2r) at (\coltwo + 1.5, -2.1 * \yoffset);
    \coordinate (arrow2) at (\colone + 0.75, -1.9 * \yoffset);

    \coordinate (arrow3) at (\colone + 0.7, -2.7 * \yoffset);

    \coordinate (left4l) at (\colone, -4 * \yoffset);
    \coordinate (right4l) at (\colone + 1.5, -4 * \yoffset);
    \coordinate (label4r) at (\coltwo - 3/4, -4 * \yoffset);
    \coordinate (left4r) at (\coltwo, -4 * \yoffset);
    \coordinate (right4r) at (\coltwo + 1.5, -4 * \yoffset);
    \coordinate (arrow4) at (\colone + 0.75, -3.75 * \yoffset);

    \begin{scope}[scale=1]
        \begin{scope}[scale=1/2, shift={(left1l)}]
            \node [node, fill=plotgrey] at (-0.5, 1) (lhs) {};
            \node [node] at (-0.5, 0) (lhs1) {};
        \end{scope}
        \node [] at (arrow1) () {$\xRightarrow{\times 1}$};
        \begin{scope}[scale=1/2, shift={(right1l)}]
            \node [node, fill=plotgrey] at (0.5, 1) (rhs) {};
            \node [node] at (0.5, 0) (rhs1) {};
            \draw [edge] (rhs) -- (rhs1);
        \end{scope}
    \end{scope}

    \begin{scope}[scale=1]
        \begin{scope}[scale=1/2, shift={(left2l)}]
            \node [node] at (-0.5, 1) (lhs) {};
            \node [node] at (-0.5, 0) (lhs1) {};
        \end{scope}
        \node [] at (arrow2) () {$\xRightarrow{\times 5}$};
        \begin{scope}[scale=1/2, shift={(right2l)}]
            \node [node] at (0.5, 1) (rhs) {};
            \node [node] at (0.5, 0) (rhs1) {};
            \draw [edge] (rhs) -- (rhs1);
        \end{scope}
    \end{scope}

    \node at (arrow3) {$\vdots$};

    \begin{scope}[scale=1]
        \begin{scope}[scale=1/2, shift={(left4l)}]
            \node [node, fill=plotgrey] at (-0.75, 0.5) (lhs) {};
            \node [node] at (0, 0) (lhs1) {};
            \node [node] at (0, 1) (lhs2) {};
            \draw [edge] (lhs) -- (lhs1);
            \draw [edge] (lhs) -- (lhs2);
        \end{scope}
        \node [] at (arrow4) () {$\xRightarrow{\times 3}$};
        \begin{scope}[scale=1/2, shift={(right4l)}]
            \node [node, fill=plotgrey] at (0, 0.5) (rhs) {};
            \node [node] at (0.75, 0) (rhs1) {};
            \node [node] at (0.75, 1) (rhs2) {};
            \draw [edge] (rhs) -- (rhs1);
            \draw [edge] (rhs) -- (rhs2);
            \draw [edge] (rhs1) -- (rhs2);
        \end{scope}
    \end{scope}

    \begin{pgfonlayer}{background} 
        \coordinate (sw) at (current bounding box.south west);
        \coordinate (se) at (current bounding box.south east);
        \coordinate (nw) at (current bounding box.north west);
        \coordinate (ne) at (current bounding box.north east);

        \draw [draw=plotgrey, fill=plotgrey, opacity=0.75] ([shift={(-0.2, -0.1)}]sw) rectangle ([shift={(0.2, 0.1)}]ne);
        \draw [very thick, draw=white] ([shift={(-0.21, -1)}]nw) -- ([shift={(0.21, -1)}]ne);
        \draw [very thick, draw=white, fill=white] ([shift={(-0.21, 1)}]sw) rectangle ([shift={(0.21, 1.8)}]se);
    \end{pgfonlayer}
\end{tikzpicture}
}
    ~~~~~~~~~~~~~~~~
    \scalebox{2}{\begin{tikzpicture}
    \begin{scope}[scale=1]  
        \node [node, draw=plotblack, fill=plotgrey] at (0, 0) (ego) {$v$};

        \node [node, draw=plotred, fill=plotred] at (0, 1) (b) {};
        \node [node, draw=plotorange, fill=plotorange] at (-0.7071067, 0.7071067) (c) {};
        \node [node, draw=plotyellow, fill=plotyellow] at (-1, 0) (d) {};
        \node [node, draw=plotteal, fill=plotteal] at (1, 0) (e) {};
        \node [node, draw=plotblue, fill=plotblue] at (0.7071067, -0.7071067) (f) {};
        \node [node, draw=plotgreen, fill=plotgreen] at (0.7071067, 0.7071067) (g) {};
    \end{scope}
    \begin{scope}[scale=1] 
        \draw [edge] (ego) -- (b);
        \draw [edge] (ego) -- (c);
        \draw [edge] (ego) -- (d);
        \draw [edge] (ego) -- (e);
        \draw [edge] (ego) -- (f);
        \draw [edge] (ego) -- (g);

        \draw [edge] (b) -- (c);
        \draw [edge] (b) -- (d);
        \draw [edge] (b) -- (g);
        \draw [edge] (c) -- (d);
        \draw [edge] (e) -- (g);
    \end{scope}
\end{tikzpicture}
}
    \captionof{figure}{A sample of the temporal subgraph transitions (middle) between the time $t$ (left) and time $t + 1$ (right) egonets of a node $v$.}
    \label{fig:teaser}
    ~\\
{}}}{}{}
\makeatother

\begin{document}

\newcommand\teaser{%
    \centering
    \scalebox{2}{\begin{tikzpicture}
    \begin{scope}[scale=1]  
        \node [node, draw=plotblack, fill=plotgrey] at (0, 0) (ego) {$v$};

        \node [node, draw=plotred, fill=plotred] at (0, 1) (b) {};
        \node [node, draw=plotorange, fill=plotorange] at (-0.7071067, 0.7071067) (c) {};
        \node [node, draw=plotyellow, fill=plotyellow] at (-1, 0) (d) {};
        \node [node, draw=plotteal, fill=plotteal] at (1, 0) (e) {};
        \node [node, draw=plotblue, fill=plotblue] at (0.7071067, -0.7071067) (f) {};
    \end{scope}
    \begin{scope}[scale=1] 
        \draw [edge] (ego) -- (b);
        \draw [edge] (ego) -- (c);
        \draw [edge] (ego) -- (d);
        \draw [edge] (ego) -- (e);
        \draw [edge] (ego) -- (f);

        \draw [edge] (e) -- (f);
    \end{scope}
\end{tikzpicture}
}
    ~~~~~~~~~~~~~~~~
    \scalebox{1}{\tikzmath{%
    \colone = 0;
    \coltwo = 5;
    \yoffset = 1;
    \yshiftrulefour = -0.75;
}

\begin{tikzpicture}
    \coordinate (left1l) at (\colone, -1 * \yoffset);
    \coordinate (right1l) at (\colone + 1.5, -1 * \yoffset);
    \coordinate (label1r) at (\coltwo - 3/4, -1 * \yoffset);
    \coordinate (left1r) at (\coltwo, -1 * \yoffset);
    \coordinate (right1r) at (\coltwo + 1.5, -1 * \yoffset);
    \coordinate (arrow1) at (\colone + 0.75, -0.75 * \yoffset);

    \coordinate (left2l) at (\colone, -2.1 * \yoffset);
    \coordinate (right2l) at (\colone + 1.5, -2.1 * \yoffset);
    \coordinate (label2r) at (\coltwo - 3/4, -2.1 * \yoffset);
    \coordinate (left2r) at (\coltwo, -2.1 * \yoffset);
    \coordinate (right2r) at (\coltwo + 1.5, -2.1 * \yoffset);
    \coordinate (arrow2) at (\colone + 0.75, -1.9 * \yoffset);

    \coordinate (arrow3) at (\colone + 0.7, -2.7 * \yoffset);

    \coordinate (left4l) at (\colone, -4 * \yoffset);
    \coordinate (right4l) at (\colone + 1.5, -4 * \yoffset);
    \coordinate (label4r) at (\coltwo - 3/4, -4 * \yoffset);
    \coordinate (left4r) at (\coltwo, -4 * \yoffset);
    \coordinate (right4r) at (\coltwo + 1.5, -4 * \yoffset);
    \coordinate (arrow4) at (\colone + 0.75, -3.75 * \yoffset);

    \begin{scope}[scale=1]
        \begin{scope}[scale=1/2, shift={(left1l)}]
            \node [node, fill=plotgrey] at (-0.5, 1) (lhs) {};
            \node [node] at (-0.5, 0) (lhs1) {};
        \end{scope}
        \node [] at (arrow1) () {$\xRightarrow{\times 1}$};
        \begin{scope}[scale=1/2, shift={(right1l)}]
            \node [node, fill=plotgrey] at (0.5, 1) (rhs) {};
            \node [node] at (0.5, 0) (rhs1) {};
            \draw [edge] (rhs) -- (rhs1);
        \end{scope}
    \end{scope}

    \begin{scope}[scale=1]
        \begin{scope}[scale=1/2, shift={(left2l)}]
            \node [node] at (-0.5, 1) (lhs) {};
            \node [node] at (-0.5, 0) (lhs1) {};
        \end{scope}
        \node [] at (arrow2) () {$\xRightarrow{\times 5}$};
        \begin{scope}[scale=1/2, shift={(right2l)}]
            \node [node] at (0.5, 1) (rhs) {};
            \node [node] at (0.5, 0) (rhs1) {};
            \draw [edge] (rhs) -- (rhs1);
        \end{scope}
    \end{scope}

    \node at (arrow3) {$\vdots$};

    \begin{scope}[scale=1]
        \begin{scope}[scale=1/2, shift={(left4l)}]
            \node [node, fill=plotgrey] at (-0.75, 0.5) (lhs) {};
            \node [node] at (0, 0) (lhs1) {};
            \node [node] at (0, 1) (lhs2) {};
            \draw [edge] (lhs) -- (lhs1);
            \draw [edge] (lhs) -- (lhs2);
        \end{scope}
        \node [] at (arrow4) () {$\xRightarrow{\times 3}$};
        \begin{scope}[scale=1/2, shift={(right4l)}]
            \node [node, fill=plotgrey] at (0, 0.5) (rhs) {};
            \node [node] at (0.75, 0) (rhs1) {};
            \node [node] at (0.75, 1) (rhs2) {};
            \draw [edge] (rhs) -- (rhs1);
            \draw [edge] (rhs) -- (rhs2);
            \draw [edge] (rhs1) -- (rhs2);
        \end{scope}
    \end{scope}

    \begin{pgfonlayer}{background} 
        \coordinate (sw) at (current bounding box.south west);
        \coordinate (se) at (current bounding box.south east);
        \coordinate (nw) at (current bounding box.north west);
        \coordinate (ne) at (current bounding box.north east);

        \draw [draw=plotgrey, fill=plotgrey, opacity=0.75] ([shift={(-0.2, -0.1)}]sw) rectangle ([shift={(0.2, 0.1)}]ne);
        \draw [very thick, draw=white] ([shift={(-0.21, -1)}]nw) -- ([shift={(0.21, -1)}]ne);
        \draw [very thick, draw=white, fill=white] ([shift={(-0.21, 1)}]sw) rectangle ([shift={(0.21, 1.8)}]se);
    \end{pgfonlayer}
\end{tikzpicture}
}
    ~~~~~~~~~~~~~~~~
    \scalebox{2}{\begin{tikzpicture}
    \begin{scope}[scale=1]  
        \node [node, draw=plotblack, fill=plotgrey] at (0, 0) (ego) {$v$};

        \node [node, draw=plotred, fill=plotred] at (0, 1) (b) {};
        \node [node, draw=plotorange, fill=plotorange] at (-0.7071067, 0.7071067) (c) {};
        \node [node, draw=plotyellow, fill=plotyellow] at (-1, 0) (d) {};
        \node [node, draw=plotteal, fill=plotteal] at (1, 0) (e) {};
        \node [node, draw=plotblue, fill=plotblue] at (0.7071067, -0.7071067) (f) {};
        \node [node, draw=plotgreen, fill=plotgreen] at (0.7071067, 0.7071067) (g) {};
    \end{scope}
    \begin{scope}[scale=1] 
        \draw [edge] (ego) -- (b);
        \draw [edge] (ego) -- (c);
        \draw [edge] (ego) -- (d);
        \draw [edge] (ego) -- (e);
        \draw [edge] (ego) -- (f);
        \draw [edge] (ego) -- (g);

        \draw [edge] (b) -- (c);
        \draw [edge] (b) -- (d);
        \draw [edge] (b) -- (g);
        \draw [edge] (c) -- (d);
        \draw [edge] (e) -- (g);
    \end{scope}
\end{tikzpicture}
}
    
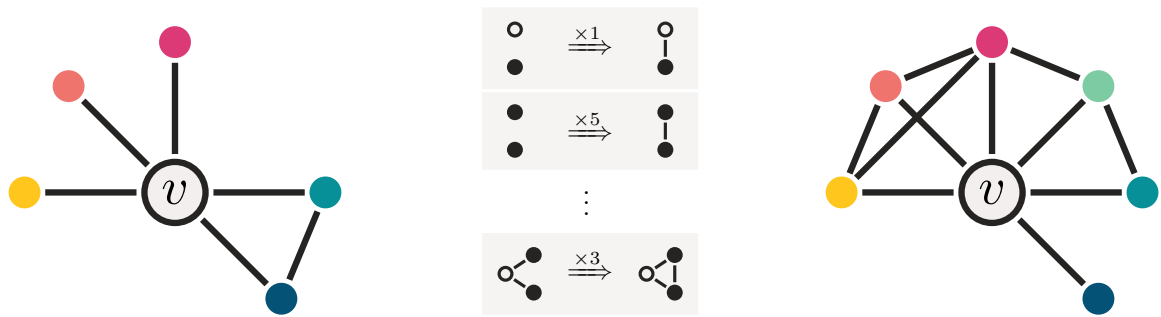
\captionof{figure}{A sample of the temporal subgraph transitions (middle) between the time $t$ (left) and time $t + 1$ (right) egonets of a node $v$.}
    \label{fig:teaser}
    ~\\
}
\title{Temporal Egonet Subgraph Transitions}

\author{Daniel Gonzalez Cedre\\
\textit{\small University of Notre Dame}\\
\textit{\small Notre Dame, USA}
\and
Sophia Abraham\\
\textit{\small University of Notre Dame}\\
\textit{\small Notre Dame, USA}
\and
Lucas Parzianello\\
\textit{\small University of Notre Dame}\\
\textit{\small Notre Dame, USA}
\and
Eric Tsai\\
\textit{\small University of Notre Dame}\\
\textit{\small Notre Dame, USA}
}
\maketitle

\begin{abstract}
    How do we summarize dynamic behavioral interactions?
    We introduce a possible node-embedding-based solution to this question: temporal egonet subgraph transitions.%
    \footnote{Source code at {\scriptsize\url{https://github.com/lucaspar/cbm-project}}.}
\end{abstract}

\section{Introduction}
\label{sec:intro}

The problem of anomaly detection in graphs has a multitude of real-world applications in many fields.
An anomaly may be a rare item, event, or entity that does not fit with the more `typical' trends in a set of data - i.e. an outlier.
Examples include bank fraud in a set of financial transactions, typos in a text, intrusions in networks, `bot' users in social media, or a sudden drop in sales caused by a global pandemic.
In this paper, we propose \textbf{T}emporal \textbf{E}gonet \textbf{S}ubgraph \textbf{T}ransitions (TEST): a comprehensive way of summarizing the behavioral patterns of nodes in a network.

We represent interactions between users over time as a sequence of graphs, representing the evolution of interactions between a set of nodes over time.
In this framework, we are interested in modeling the behavior individual nodes exhibit in order to distinguish different patterns of interaction.
To that end, TEST creates a vector representation of each node using the changes in its neighborhood---or \emph{egonet}---over time.
These vector representations can then be used for traditional tasks, such as clustering and classification, depending on the application and data quality.

TEST is flexible enough to lend itself to a variety of standard tasks and application domains because it does not depend on labeled data or any kind of iterative learning procedure.
Nodes' vector representations are computed purely based on how their egonets change over time.
Once in the embedding space, the data can be analyzed from various perspectives.
For example, given labeled data, the latent representations can be used for classification tasks such as bot-detection on social networks.
However, in the absence of ground truth regarding nodes, the latent representations can still be analyzed to find \emph{communities}: clusters of points in the embedding space signalling nodes with similar temporal behavior.

Despite its flexibility, TEST makes some important assumptions about the data.
Not only should the data be graphical, it should be a discrete sequence $G_1, G_2, \dots G_T$, where the events described by $G_{t}$ precede those in $G_{t+1}$.
This means that datasets presented as streams of timestamped edges must be \emph{discretized}.
If the discretization process is too coarse, signals in the data will be lost;
however, if the discretization is too fine, the model will not significantly compress the data.
While some datasets and application domains are concerned with clustering or classifying edges---as is the case in, for example, detecting fraudulent transactions on financial networks---our focus in this paper is on characterizing nodes' behavior.

\section{Related Work}
\label{sec:related}

Diverse techniques exists for anomaly detection in literature. Approaches include density-based techniques like k-nearest neighbor \cite{knorr2000distance}, local outlier factor \cite{breunig2000lof}, isolation forests \cite{liu2012isolation} and other variations \cite{schubert2014local}. For higher dimensional data - subspace \cite{kriegel2009outlier}, correlation \cite{kriegel2012outlier},  single class support vector machines \cite{platt1999estimating} and tensor based methods \cite{fanaee2016tensor} have been proposed. Neural network approaches include replicators \cite{hawkins2002outlier}, Bayesian Networks \cite{platt1999estimating}, hidden markov models \cite{platt1999estimating}. Additionally, cluster analysis based \cite{campello2015hierarchical}, fuzzy logic and ensemble techniques with feature bagging \cite{nguyen2010mining} and score normalization \cite{schubert2012evaluation} have been applied. 

Although these methodologies have minimal systematic advantages across data sets and parameters, many are applied in the realm of static graphs. 
To narrow the scope in relation to our proposed method regarding anomaly detection algorithms for dynamic graph networks a deeper exploration will be applied to network embedding and streaming anomaly approaches. 

\subsection{Streaming Anomaly Detection}
Streaming anomaly detection attempts to detect suspicious or uncharacteristic behavior in continuous-time streams of relational data.
\subsubsection{Streaming Graphs}
Streaming graphs can be applied to anomalous node detection like dynamic tensor analysis (DTA) \cite{sun2006beyond} which approximates an adjacency matrix for a graph at time \textit{t} with incremental matrix factorization and utilizes a high reconstruction error to.  Anomalous subgraph detection is illustrated in methods like \cite{shin2018patterns} which uses \textit{k-core} which is the maximal subgraph in which all vertices have degree at least \textit{k} and use patterns related to \textit{k-core} to find anomalous subgraphs. Streaming graphs can also be utilized for anomalous event detection such as changes in first and second derivatives in PageRank \cite{yoon2019fast}.
\subsubsection{Streaming Edges}
Streaming anomaly detection can also be applied to streaming edges. Anomalous nodes can be detected using methods like \cite{yu2013anomalous} which uses an incremental eigenvector update algorithm based in von Mises interations and discover hotspots of local changes in dynamics streams.
    SedanSpot \cite{eswaran2018sedanspot} utilizes streaming edges to detect anomalous edges by exploiting edges that connect part of the graph that are sparsely connected and finding where bursts of activity indicative of anomalous behavior.

\subsection{Network Embedding}
Network embedding approaches consist of learning low-dimensional feature representation of the nodes or links within a network. Many advances \cite{chang2015heterogeneous,levy2014neural,wang2016structural, yu2013embedding} in the network embedding space present new ways to learn representations for networks which can be used to detect anomalies. Feature learning strategies for extracting network embedding have been widely used in language models such as Skip-gram \cite{mikolov2013distributed} in which the the defining neighborhood attributes for words in a sentence are preserved. In a similar manner, many popular methodologies such as DeepWalk \cite{perozzi2014deepwalk} and Node2vec \cite{grover2016node2vec} use sequences of vertices from the graphs and learn the representation of these vertices by maximizing the preservation of the structure inherent to a neighborhood of vertices in the network. A large portion of work has also been dedicated to representation learning in other manners for graph network architectures \cite{tian2014learning, wang2016structural, zhou2018dynamic} and compact representations of the graph have been used to find anomalies within the network \cite{manzoor2016fast}. 

These methods have also been specifically applied for anomaly detection in dynamic temporal graphs. Specifically, NetWalk \cite{yu2018netwalk} learns network representations which are dynamically updated as the network evolves. This is done by by learning the latent network representations by extracted sequences of vertices from the graph, otherwise termed as "walks" from the initial network and then extract deep representations by minimizing the pairwise distance of the vertices when encoding the vector representations and adding global regularization by using a deep autoencoder reconstruction error. NetWalk is updated over dynamic changes with reservoir sampling strategy and a dynamic clustering model is used to find anomalies. 

In \cite{hibshman2021sst}, dynamic graph evolution is modeled with subgraph to subgraph transitions (SST) by fitting linear SVM models to SST count vectors. This can be used for link prediction of static, temporal, directed and undirected graphs while providing interpretable results. The idea of counting SSTs inspires our proposed methodology, however, instead of utilizing an edge based approach as this paper did, we propose a neighborhood based approach for detecting anomalies based on clusters (\textit{cf.} \autoref{sec:model}).

\section{Data Sources}
\label{sec:data-sources}

We are focused on modeling behavior over time from graph-structured data, so we need datasets with the following attributes:
\begin{enumerate}
    \item Nodes with persistent IDs.
    \item Interactions (directed or undirected) between nodes.
    \item Timestamps for the interactions.
\end{enumerate}

\subsection{Available Datasets}\label{sec:datasets}
Although we initially wanted to tackle node classification problems using TET, we ran into problems finding datasets which had both timestamped edges and ground-truth labels for the nodes.
Most of the datasets with ground-truth we found had labels for the edges instead.
Therefore, we are considering unsupervised tasks like clustering for testing the quality of our node embeddings.
We describe some of the datasets we found below.



\subsubsection{Enron Emails\footnote{\texorpdfstring{\scriptsize\url{http://odds.cs.stonybrook.edu/enroninc-dataset/}}{hi}}}
The nodes of this dataset represent email addresses and directed edges depict sent/received relations. Enron email network dataset contains a total of 80, 884 nodes and 700 timestamp in total.

\subsubsection{College Messages\footnote{\texorpdfstring{\scriptsize\url{https://snap.stanford.edu/data/CollegeMsg.html}}{hi}}}

This is a dataset containing private messages at an online social network at the University of California, Irvine. Here an edge (u, v, t) means that during time t, an user u sent a message to an user v.


\subsubsection{EU Emails\footnote{\texorpdfstring{\scriptsize\url{https://snap.stanford.edu/data/email-Eu-core-temporal.html}}{hi}}}

This dataset contains 986 nodes and 332,334 temporal edges, with the timestamp of 803 in total.

\subsection{Synthetic Dataset}
\label{sec:synth-dataset}

In order to test our model's ability to properly distinguish between nodes with different temporal behavior in the embedding space, we need a dataset which:

\begin{enumerate}
    \item represents interactions between different objects,
    \item is temporal (i.e., interactions occur at known points in time, which may repeat),
    \item has ground truth labels on the \emph{nodes} indicating whether those nodes are anomalous or not.
\end{enumerate}

There are many public temporal graph datasets that immediately satisfy the first two requirements.

Among those, there are a few that also include some ground truth information regarding authentic/suspicious/anomalous edges in the graph.
However, since we are attempting to characterize \emph{nodes} as anomalous or not, edge-centric ground truth labels are unhelpful.
Since a dataset satisfying all three criteria is unavailable to us, we could go out into the world and collect, mine, or survey a sufficiently high-quality dataset on our own.

We elected to not pursue this idea not only because of time constraints, but also because building a real-world dataset satisfying those three criteria by hand would require a level of description, analysis, and methodological detail that might be more appropriate for its own paper.

Therefore, we elected to generate a synthetic temporal graph with ground truth node labels.

The data consists of a discrete sequence of five graphs $G_1, G_2, \dots G_5$ on a shared set of $500$ nodes $V$, representing five snapshots of interactions between the nodes in $V$.

The nodes are split between \emph{authentic} and \emph{anomalous} nodes, labels which persist through the different graphs in the sequence.
For each time $t \in {0, \dots 4}$ authentic nodes' edges are generated by an Erdos-Renyi random process \cite{erdos_random_1959};

specifically, between any two authentic nodes $u$ and $v$, an edge $e = \left\{u, v\right\}$ exists with probability $p \in [0, 1]$.
The anomalous nodes are connected together in the following way:
for $t \in \left\{0, 2, 4\right\}$, all of the anomalous nodes form the empty subgraph; for $t \in \left\{1, 3, 5\right\}$, the anomalous nodes form a clique.


The free parameters for generating this dataset are the connection probability $p$ for the authentic nodes, the total number of nodes $n$, and the percentage of the total nodes that is designated as anomalous.
In the future, we can experiment with different ways of connecting the anomalous nodes together to create more-varied temporal patterns.

\section{Description of the Model}
\label{sec:model}

\begin{table*}[th!]
    \centering
    \scalebox{0.85}{
        \begin{tabular}{r|ccc|ccc|ccc|ccc|c}
            \multicolumn{14}{c}{\textbf{Synthetic dataset: $p = 0.0025$, $a = 0.05$}}\\[1ex]
            \multicolumn{1}{c}{\textbf{}} & \multicolumn{3}{c}{\textbf{Egonet}}    & \multicolumn{3}{c}{\textbf{Node2Vec}} & \multicolumn{3}{c}{\textbf{Spectral Clustering}} & \multicolumn{3}{c}{\textbf{Deepwalk}} & \\
                & Prec. & Recall & F1-Score & Prec. & Recall & F1-Score & Prec. & Recall & F1-Score & Prec. & Recall & F1-Score & Data Points \\
            Anomaly & 1.00 & 1.00 & 1.00 & 0.00 & 0.00 & 0.00 & 0.01 & 0.04 & 0.02 & 0.27 & 0.56 & 0.36 & 25 \\
            Normal & 1.00 & 1.00 & 1.00 & 0.95 & 0.95 & 0.95 & 0.94 & 0.82 & 0.88 & 0.98 & 0.92 & 0.95 & 475 \\
            Accuracy & & & 1.00 & & & 0.90 & & & 0.78 & & & 0.90 & 500                   
        \end{tabular}
    }
\caption{Where $p$ is is the connection probability for the authentic nodes, and a is the percentage of the anomalous nodes.} \label{tab:my-table}
\end{table*}

The basic idea behind TET is that changes in a node's egonet can be summarized by counting \emph{subgraph transitions}.
A subgraph transition for a node $v$ at a given time transition $(t, t+1)$ is defined as a pair of graphs $(H_t, H_{t+1})$ on a shared node set $V$ such that:

\begin{itemize}
    \item[$\cdot$] $H_t$ is a subgraph of the adjacency neighborhood of $v$ at time $t$ if any missing nodes from the neighborhood at time $t+1$ are included
    \item[$\cdot$] $H_{t+1}$ is a subgraph of the adjacency neighborhood of $v$ at time $t+1$ if any missing nodes from the neighborhood at time $t$ are included
    \item[$\cdot$] $H_t$ and $H_{t+1}$ are not isomorphic.
\end{itemize}

In the interest of computational tractability, we only consider subgraph transitions up to a certain number of vertices---for this paper, we only consider subgraph transitions on at most $N = 3$ nodes.

By enumerating and canonically ordering all of the subgraph transitions in question, we can construct a vector of \emph{subgraph transition counts} for the node $v$ during the time hop from $t$ to $t+1$.
This subgraph transition count vector summarizes the change in $v$'s behavior from the previous time step to the next time step.

Performing this for each time transition results in a sequence of count vectors $\vec{v}_1, \vec{v}_2, \dots \vec{v}_{T-1} \in \mathbb{R}^d$, where $d$ is the number of subgraph transitions enumerated (note: $d$ may be \emph{very} large, even for small $N$).
We can then condense these counts into a single vector representation for $v$ by applying some element-wise \emph{aggregation function} (e.g., mean, sum, min, max).
In this paper, we aggregate the count vectors by averaging each of the transition counts over all of the time steps, producing $\vec{v} = \frac{1}{T-1}\sum_{t=1}^{T-1}\vec{v}_t \in \mathbb{R}^d$ as the embedding vector for the node in question.
The model is described in detail in \autoref{fig:pseudocode}.

\begin{algorithm}[h]
\SetAlgoLined
\KwIn{List of subgraph transitions $\mathcal{T}$\\
\phantom{eeeeeii} Temporal graph sequence, $G_1, \dots G_T$\\
\phantom{eeeeeii} Set of nodes $V$\\}
\KwOut{Set of embeddings $\left\{\vec{c}_v ~|~ v \in V\right\}$}
\For{$v$ in $V$}{
    \For{$t \in \left\{1, \dots T-1\right\}$}{
        Initialize $\vec{c}_{v, t} = (0, \dots 0) \in \mathbb{R}^{|\mathcal{T}|}$\;
        Let $G_{t}(v)$ be the egonet of $v$ at time $t$\;
        Let $G_{t+1}(v)$ be the egonet of $v$ at time $t+1$\;
        Union the vertex sets of $G_{t}(v)$ and $G_{t+1}(v)$\;
        \For{$(L, R)_i \in \mathcal{T}$}{
            \For{$G \lhd G_{t}(v)$ isomorphic to $L$}{
                \For{$H \lhd G_{t+1}(v)$ isomorphic to $R$}{
                    \eIf{$G$ and $H$ share the same nodes}{
                        Increment index $i$ of $\vec{c}_{v, t}$\;
                    }{
                        Continue\;
                    }
                }
            }
        }
    }
Compute $\vec{c}_v = \frac{1}{T-1}\sum_{t=1}^{T-1}\vec{c}_{v, t}$\;
{\bf yield} $\vec{c}_v$\;
}
\caption{\label{fig:pseudocode}how to embed temporal egonets}
\end{algorithm}

%


\section{Evaluation}
\label{sec:evaluation}

\subsection{Experiments Setup}

In the case that we find a properly labeled dataset to evaluate TET, we can use more commonly found metrics to compare our results in an anomaly-detection problem. We probably will not use accuracy of classification directly, however. That is because anomaly datasets in general are heavily skewed towards the normal behavior, thus, any classifier that labeled them all as normal would achieve an accuracy close to 100\%. Thus, we focus our report in the F1-score, precision, and recall achieved in the baseline executions and compare them against Egonet. The experiments shown in Table 1 shown here were carried out on the synthetic dataset described in \autoref{sec:synth-dataset}.

\subsection{Baselines}
In order to assess the quality of our method, we will compare against competing methods and apply the same unsupervised learning methods later on.

\begin{itemize}

    \item[$\cdot$] \textbf{Spectral Clustering~\cite{liu2013spectral}}: is closely related to nonlinear dimension reduction, and dimension reduction techniques such as locally-linear embedding can be used to reduce errors from noise. By calculating the Laplacian of the graph and obtaining the first eigenvectors and corresponding eigenvalues of the Laplacian, this method output embedding nodes. Spectral Clustering has no assumption on the shapes of the clusters and can thus model complex scenarios. 
    
    \item[$\cdot$] \textbf{DeepWalk~\cite{perozzi2014deepwalk}}: generalizes on language modeling from sequences of words to graphs by using local information from sequences of vertices(Random Walk and SkipGram) in the graph and learning representations by treating them equivalently to sentences. Here the input is an edge list file and the output of DeepWalk is 64 dimensions by default. In order to visualize it, the output is manually set to 2 dimensions. 
    
    \item[$\cdot$] \textbf{Node2Vec~\cite{grover2016node2vec}}: is a useful method for extracting the continuous-feature representations for nodes in graph networks. By using a biased random walk procedure, Node2Vec maps of nodes to a lower-dimensional space of features while preserving their locality.
    
\end{itemize}

The baselines mentioned above do not take dynamic graphs into account without a remodeling of the input data and multiple executions, making it a challenging task to represent the evolution of the graph over time. In our case temporal graphs into account, which significantly lowers to their implementation complexity. All of the output of embedding methods (including the output of Egonet) are later fed forward to the same pipeline of DBSCAN for outlier detection. The results are displayed in Figures 1, 2, and 3. The images also point the frequency distribution among the classes found by DBSCAN. As opposed to k-means, DBSCAN attempts to find the optimal number of classes, so the number of clusters is highly dependent on the embedding dataset.

\begin{figure}[hb]
    \centering
    \includegraphics[width=\linewidth]{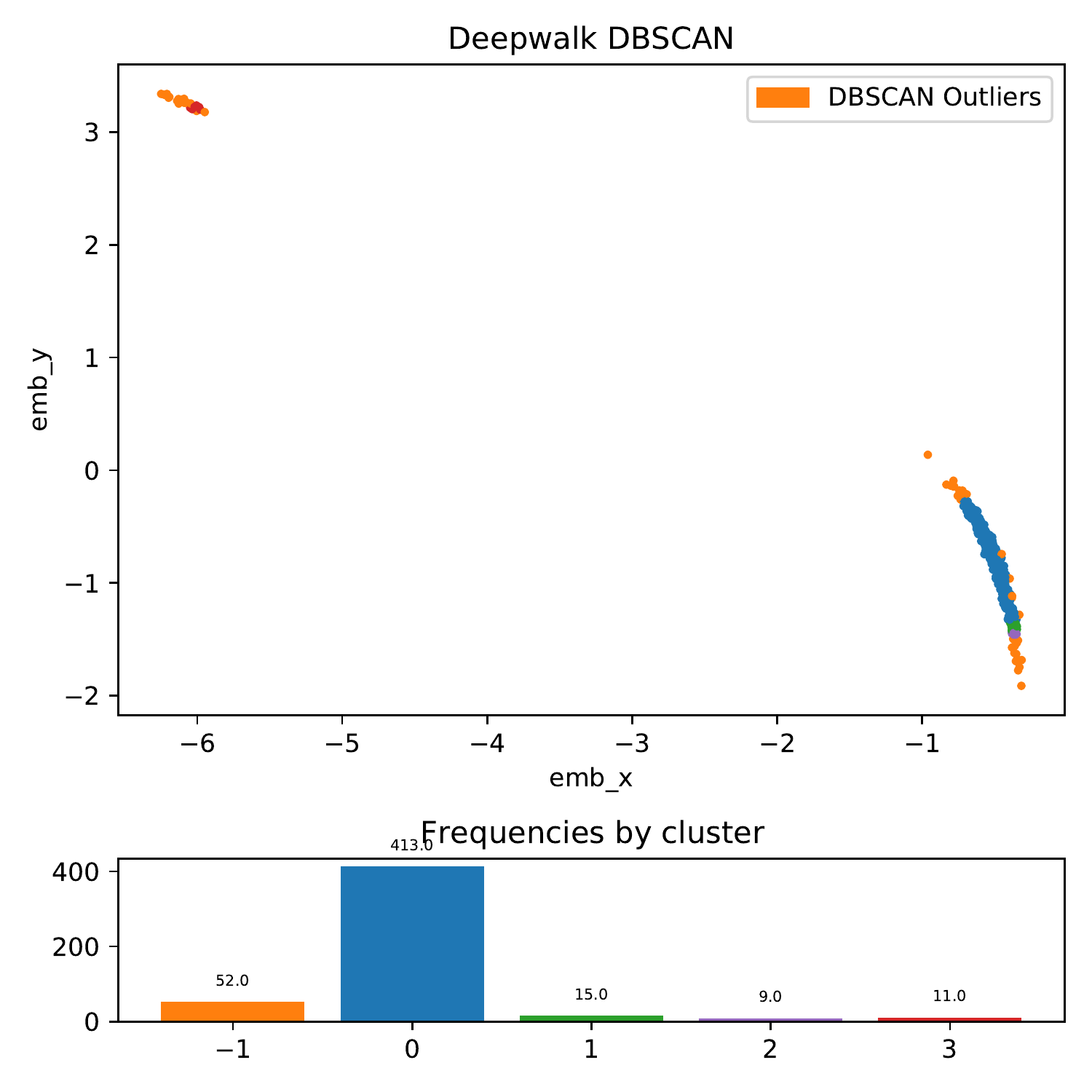}
    \caption{Results of the clustering after DeepWalk node embedding on the synthetically generated dataset.}
\end{figure}

\begin{figure}[hb]
    \centering
    \includegraphics[width=\linewidth]{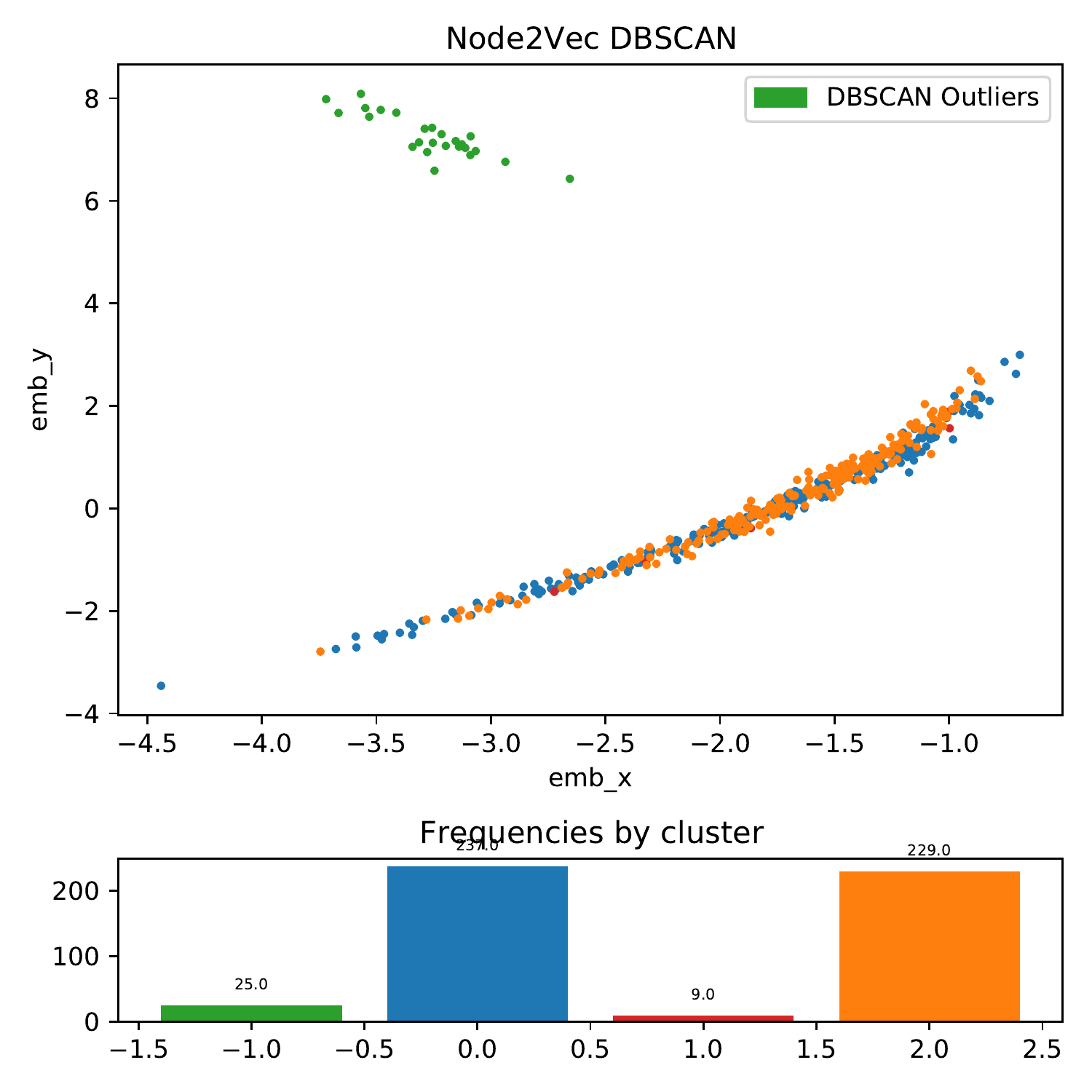}
    \caption{Results of the clustering after Node2Vec node embedding on the synthetically generated dataset.}
\end{figure}

\begin{figure}[hb]
    \centering
    \includegraphics[width=\linewidth]{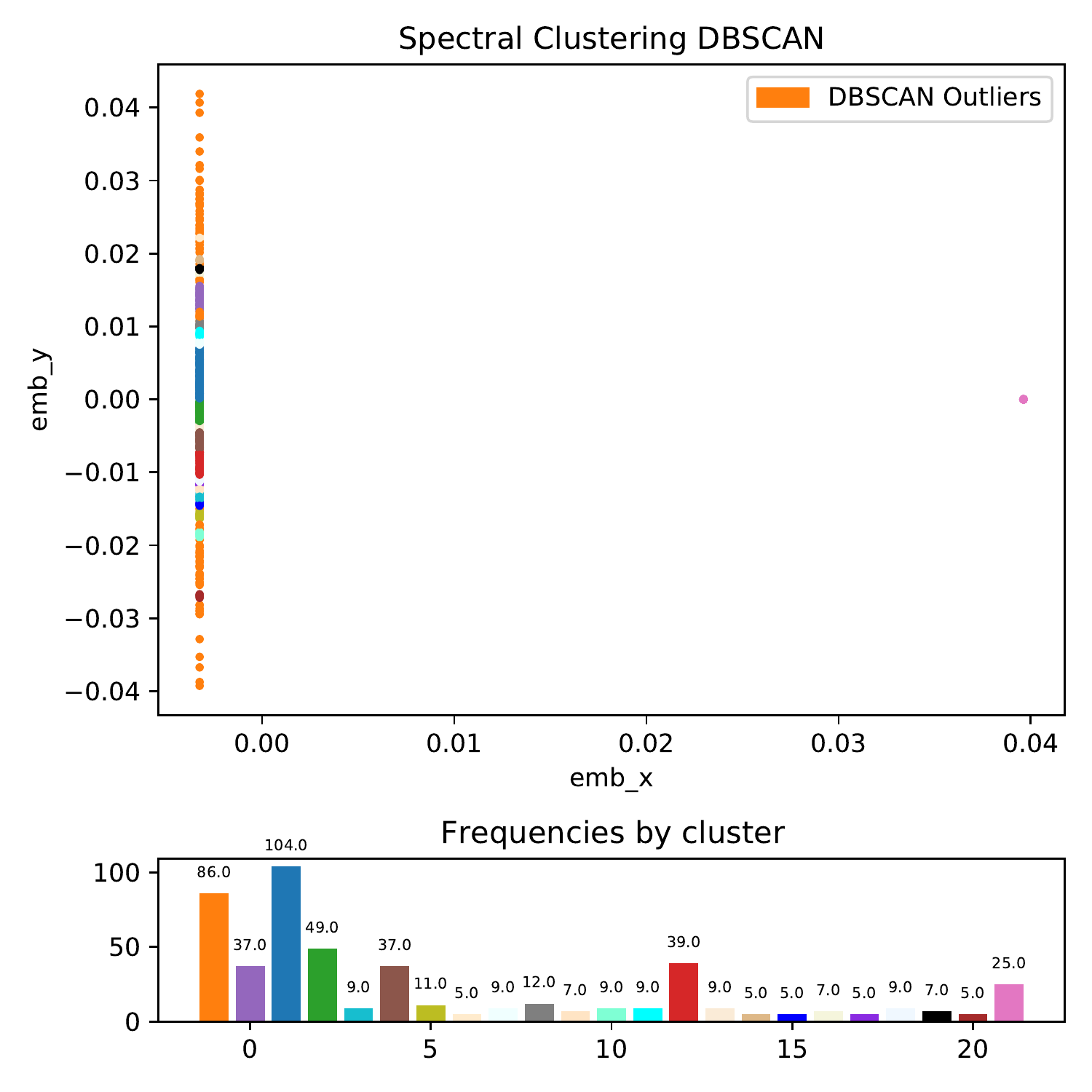}
    \caption{Results of the clustering after Spectral Clustering node embedding on the synthetically generated dataset.}
\end{figure}

\subsection{Future Work}
In addition to the clustering techniques mentioned above, the Extreme Value Machine (EVM) \cite{rudd2017extreme} could be used to classify extracted embeddings.  

The EVM is a scalable nonlinear classifier which supports open-set classification, where the models are assumed to have incomplete knowledge of their operating context and can reject inputs that are beyond the support of the training set. The EVM relies on a strong feature representation and every represented sample in the feature representation becomes a point. EVM utilizes bins which groups all the points in their feature representation by their correspondent label. These bins are utilized to create a 1 vs. rest EVM classifier for each known class or in our case each of the known clusters. The EVM generates a classifier where a Weibull distribution is fit on the data for each known class and is made to avoid the negative data points (unknown classes) or outliers for our application.  This is repeated for all the known classes/clusters.  When a coordinate point which is a sample encoded by its feature representation as a vector is provided to the EVM, the points are mapped to the feature space as a probability of belonging to a representative class, or if it falls below a threshold of inclusion, labeled as an outlier. 

The analysis of our approach for extracting embeddings and clustering with DBSCAN proved to be successful on the synthetically generated data, however, real world data may be more complex and difficult to detect outliers in. Thus, using an approach like the EVM and forming distributions that take into account the input space distance distributions may provide strong understanding of the nature of the data and prove to be an effective means of capturing anomalous nodes based off the embeddings.

{\small
\bibliographystyle{plain}
\bibliography{refs}
}

\end{document}